\pdfoutput=1 
\documentclass[screen,dvipsnames,nonacm,acmlarge]{acmart}

\copyrightyear{2022}
\setcopyright{none}
\acmYear{}
\acmConference{}{}{}
\acmPrice{}
\acmDOI{}
\acmISBN{}

\usepackage{array,multirow}
\usepackage{bbold}
 
\usepackage{adjustbox,amsmath,amsfonts,cite,float,graphicx,hyperref,morefloats,placeins,tabularx,color,lipsum,subfigure,physics}
\usepackage[font=small,skip=0pt]{caption}

\DeclareMathOperator\Dual{Du}

\begin{document}

\title{Novel algorithm for the computation of group and energy velocities of Lamb waves}

\author{Adil Han Orta}
\email{adilhan.orta@kuleuven.be}
\affiliation{%
    \department{Department of Physics}
    \institution{KU Leuven Campus Kortrijk--Kulak}
    \streetaddress{Etienne Sabbelaan 53 bus 7657}
    \postcode{8500}
    \city{Kortrijk}
    \country{Belgium}
}
\author{Martin Roelfs}
\orcid{0000-0002-8646-7693}
\email{martin.roelfs@kuleuven.be}
\affiliation{%
    \department{Department of Physics}
    \institution{KU Leuven Campus Kortrijk--Kulak}
    \streetaddress{Etienne Sabbelaan 53 bus 7657}
    \postcode{8500}
    \city{Kortrijk}
    \country{Belgium}
}
\author{Koen Van Den Abeele}
\email{koen.vandenabeele@kuleuven.be}
\affiliation{%
    \department{Department of Physics}
    \institution{KU Leuven Campus Kortrijk--Kulak}
    \streetaddress{Etienne Sabbelaan 53 bus 7657}
    \postcode{8500}
    \city{Kortrijk}
    \country{Belgium}
}

\begin{abstract}
A new solution strategy for quadratic eigenvalue problems, and the derivatives of the eigenvalues, is proposed, by combining the generalized reduction method with dual numbers.
To demonstrate the method, we use the quadratic eigenvalue problem encountered in the semi-analytical finite element method (SAFE) as a guiding example.
The SAFE method is designed to calculate the spectrum of Lamb wave phase, group and energy velocities in (visco)elastic orthotropic media, over a wide frequency range.
It was found that the new approach essentially doubles the computational speed and efficiency, without sacrificing accuracy.

\end{abstract}


\maketitle
\renewcommand{\shortauthors}{Orta, Roelfs and Van Den Abeele}

\section{Introduction}
Solving quadratic eigenvalue problems commonly leads to numerical instabilities, and is computationally inefficient. In the literature, there are certain numerical solution strategies to linearize the polynomial eigenvalue problems to increase the computational efficiency, without sacrificing accuracy \citep{datta2010numerical}. However, in certain applications, the derivatives of the eigenvalues are also required. In order to simultaneously obtain the eigenvalues and their derivatives, a novel solution strategy is proposed: first the eigenvalue problem is linearized using the Generalized Reduction method, after which dual numbers are used to perform Automatic Differentiation. We refer to this method as GRAD.

As our guiding example, we consider simulations of guided waves using the semi analytical finite element (SAFE) approach.
Guided waves are generally employed for SHM and ultrasonic inspection applications, owing to the fact that these types of waves can travel over long distances \citep{segers2020probing}. As such, large areas can be inspected with a limited number of sensors \citep{wang2018sparse}. One of the most common methods to simulate guided wave propagation in solid plates, bars or tubes, is the SAFE approach, because of its accuracy and robustness to compute the Lamb wave mode spectra \citep{bartoli2006modeling,mazzotti2014computation}. 
However, for inverse problem applications, such as the (visco)elastic parameter characterization of composite laminates, or for quasi-real time structural health monitoring applications, 
the total computational time becomes dominated by the evaluation speed of the forward model. Therefore fast and accurate evaluation of the forward model is important.
In the present study, GRAD is implemented to calculate phase, group and energy velocities of Lamb waves in solid plates. It is shown that the computational speed and efficiency is greatly enhanced by utilizing GRAD.

\section{Model}
As a guiding example for the implementation of GRAD, the quadratic eigenvalue problem of the SAFE method is considered; although the method can be extended to higher-order polynomial eigenvalue problems, or other quadratic eigenvalue problems such as Legendre polynomials \citep{bou2013legendre}, or higher order shear deformation theories \citep{orta2021modelling} as well.
The SAFE method, and its accompanying parameters, are well defined in the literature \citep{bartoli2006modeling,treyssede2008elastic}. The quadratic eigenvalue problem of the SAFE method can generally be expressed as \citep{treyssede2008elastic} 
    \begin{equation}
         \bqty{ K_3 k^2 + (K_2-K_2^T) k i + (K_1 - M \omega^2) }u(x,t) = 0,
        \label{eq:poly_eig}
    \end{equation}
where $u(x, t) = u_0 \exp[ik x-i\omega t]$ is a plane wave with initial displacement vector $u_0$, the $K_j$ are stiffness matrices, $M$ is mass matrix, $k$ is the wavenumber  ($k=2 \pi / \lambda$, with $\lambda$ the wavelength), and $\omega$ is the angular  frequency ($\omega= 2 \pi f$, with $f$ the frequency). In this equation, we are not exclusively interested in the solutions for $k$, which would give us the allowable phase velocities at a certain frequency and thus the dispersion curves, but also in ${\partial \omega}/{\partial k}$, the group velocity of the modes. While there are several existing techniques to solve for $k$ [6], in this paper we focus on demonstrating how these techniques can be easily extended to include the computation of ${\partial \omega}/{\partial k}$, up to the machine precision, by using dual numbers.

\subsection{Generalized reduction method}
The most straightforward solution strategy is to first linearize Eq.~\ref{eq:poly_eig} using the generalized reduction method \citep{datta2010numerical}. If the matrix $K_3$ is non-singular, we can exploit this to rewrite the quadratic eigenvalue equation as a linear system, by defining
    \begin{equation}
        \underbrace{\begin{bmatrix}
            0 & \mathbb{1}\\
            K_3^{-1} (K_1-M\omega^2) & K_3^{-1} (K_2-K_2^T)
        \end{bmatrix}}_{A(\omega)}
        \underbrace{\begin{bmatrix}
            u(x,t) \\
            i k u(x,t) \\
        \end{bmatrix}}_{v(\omega)}
        = i k \underbrace{\begin{bmatrix}
             u(x,t) \\
            i k u(x,t) \\
        \end{bmatrix}}_{v(\omega)}
        \label{eq:linearized_eigen}
    \end{equation}
As Eq. \ref{eq:poly_eig} and Eq. \ref{eq:linearized_eigen} are equivalent, the eigenvalues $k$ are identical for both. Therefore, the eigenvalues of Eq. \ref{eq:poly_eig} can be obtained by finding the eigenvalues of the matrix $A(\omega)$, which can be found using conventional methods. 
However, we are not only interested in the eigenvalues $k$, but also wish to simultaneously obtain $\partial k / \partial \omega$.
In order to calculate the latter, dual numbers are used.

\subsection{Automatic differentiation of eigenvalues}
Dual numbers rose to prominence in the realm of automatic differentiation \citep{revels2016forwardmode}. A dual number is an expression of the form $a + b \epsilon$, where $a,b \in \mathbb{C}$, and $\epsilon$ is defined to satisfy $\epsilon^2=0$.
Assuming the matrix function $A(\omega)$ is analytical in some neighborhood of $\omega$, it permits the Taylor expansion
    \begin{equation}
        A(\omega+\epsilon) = A(\omega) + \epsilon A'(\omega).
        \label{eq:taylor_series}
    \end{equation}
It follows that $\Dual [ A(\omega + \epsilon) ] = A'(\omega)$, where $\Dual$ selects the dual part.
Assuming the matrix $A = A(\omega)$ is diagonalizable as $A = v \Lambda v^{-1}$, where
$v = [ v_1, v_2 \ldots, v_n ]$ is a square matrix whose columns are the linearly independent eigenvectors $v_i$, and $\Lambda_{ij}(\omega) = k_i(\omega) \delta_{ij}$ is the diagonal matrix containing the eigenvalues, 
we define the vector of eigenvalues as $\vec{k}(\omega) = \text{diag}(\Lambda(\omega))$. Then the derivatives of the eigenvalues are given by \citep{lancaster1964eigenvalues}:
\begin{equation}
\vec{k}' =\dv{\vec{k}}{\omega} = \text{diag}\pqty{v^{-1}(\omega) A'(\omega) v(\omega)}.
\label{eq:der_eigenvalues}
\end{equation}
Since $A'(\omega) = \Dual[A(\omega+\epsilon)]$, these derivatives can be computed with machine precision, provided $A(\omega+\epsilon)$ can be evaluated. In general this can be done using an automatic differentiation library \citep{revels2016forwardmode}, but
in the particular case of $A(\omega)$ as defined in Eq. \ref{eq:linearized_eigen}, $A(\omega+\epsilon)$ can be evaluated analytically:
    \begin{equation}
        A'(\omega) = \begin{bmatrix}
            0 & 0\\
            2K_3^{-1} M\omega & 0 \\
        \end{bmatrix}.
    \end{equation}
This enables us to find $k_j' = {\partial k_j}/{\partial \omega}$ using Eq. \ref{eq:der_eigenvalues}, after which
the corresponding group velocity $\partial{\omega}/\partial{k_j}$ is $1 / k_j'$.

\newpage
\subsection{Calculation of energy velocity}
As previously mentioned in the literature \citep{bartoli2006modeling,treyssede2008elastic}, the group velocity definition is no longer valid in waveguides in attenuative media. For damped waves, the wavenumbers become complex, where the imaginary part carries the attenuation information. Using the group velocity definition, the derivative of the real part of the complex wavenumber yields nonphysical solutions such as infinite velocities at some frequencies. At this point, the energy velocity $V_e$ is considered the appropriate property for damped media. The general expression of the energy velocity reads \citep{treyssede2008elastic,mazzotti2012guided}:
    \begin{equation}
    V_e(\omega) =\frac{\Im(u^H 2 \omega (K_2^T+K_3 i k)u)}{\Re(u^H(K_3 k^2 + (K_2 - K_2^T)i k + K_1 + M \omega^2)u)} \label{eq:simplified_energy}
    \end{equation}
However, the eigenvectors $A(\omega)$, as defined in Eq. \ref{eq:linearized_eigen}, are $v_j~=~\smqty[ u_j & i k_j u_j ]^T$. This allows Eq. \ref{eq:simplified_energy} to be further simplified to
    \begin{align}
    V_e(\omega) = \frac{2\omega\Im\bqty{ \text{diag}(v^H A_1 v)}}{\Re \bqty{ \text{diag}(v^H A_2 v)}}, \label{eq:energy_vel}
    \end{align}
where $A_1 = \sbmqty{K_2^T & 0\\ -K_3 & 0 }$ and $A_2 = \sbmqty{K_1+M\omega^2 & 0\\ -(K_2-K_2^T) & K_3}$.
Note that the energy velocity exactly reverts to the group velocity in the case of undamped wave propagation.

\section{Results}
To validate the proposed group and energy velocity computation model, a numerical study is conducted for which the GRAD results are compared with the conventional SAFE software GUIGUW \citep{bocchini2011graphical}. The propagation characteristics of Lamb waves are examined for a homogenized purely elastic orthotropic carbon/epoxy (C/E) composite plate and a visco-elastic version of the same plate \citep{bartoli2006modeling}, with material stiffness (and viscosity) tensor components as listed in Table \ref{table:materials}. The material density of the C/E material is $\rho=1571$ kg/m$^3$ and the plate thickness is 1 mm.

\begin{figure}[htb]
\subfigure[ ]{\includegraphics[trim={2.5cm 1cm 2.4cm 2cm},clip=true, scale=0.25, angle=0]{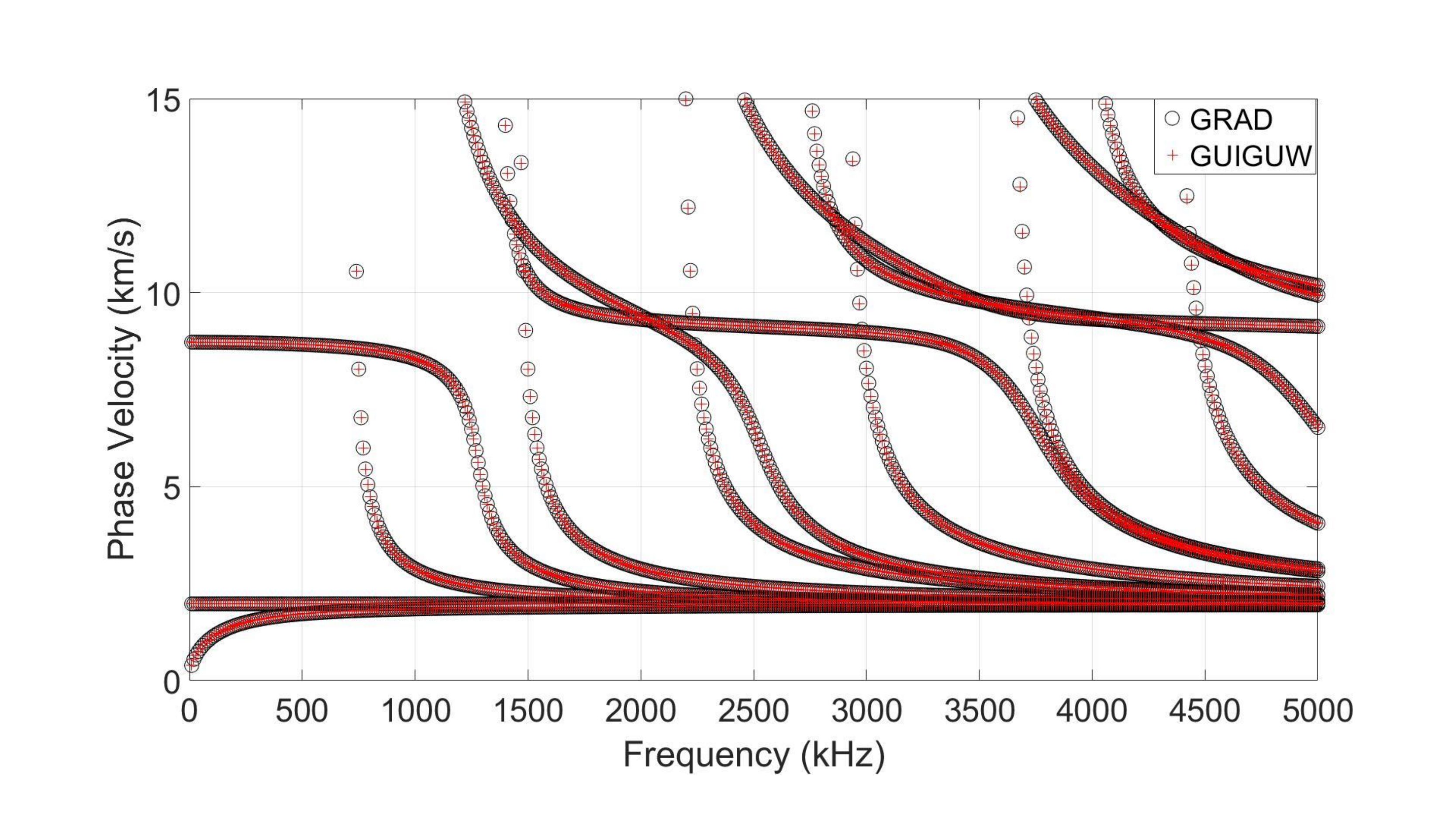}}
\subfigure[ ]{\includegraphics[trim={2.5cm 1cm 2.4cm 2cm},clip=true, scale=0.25, angle=0]{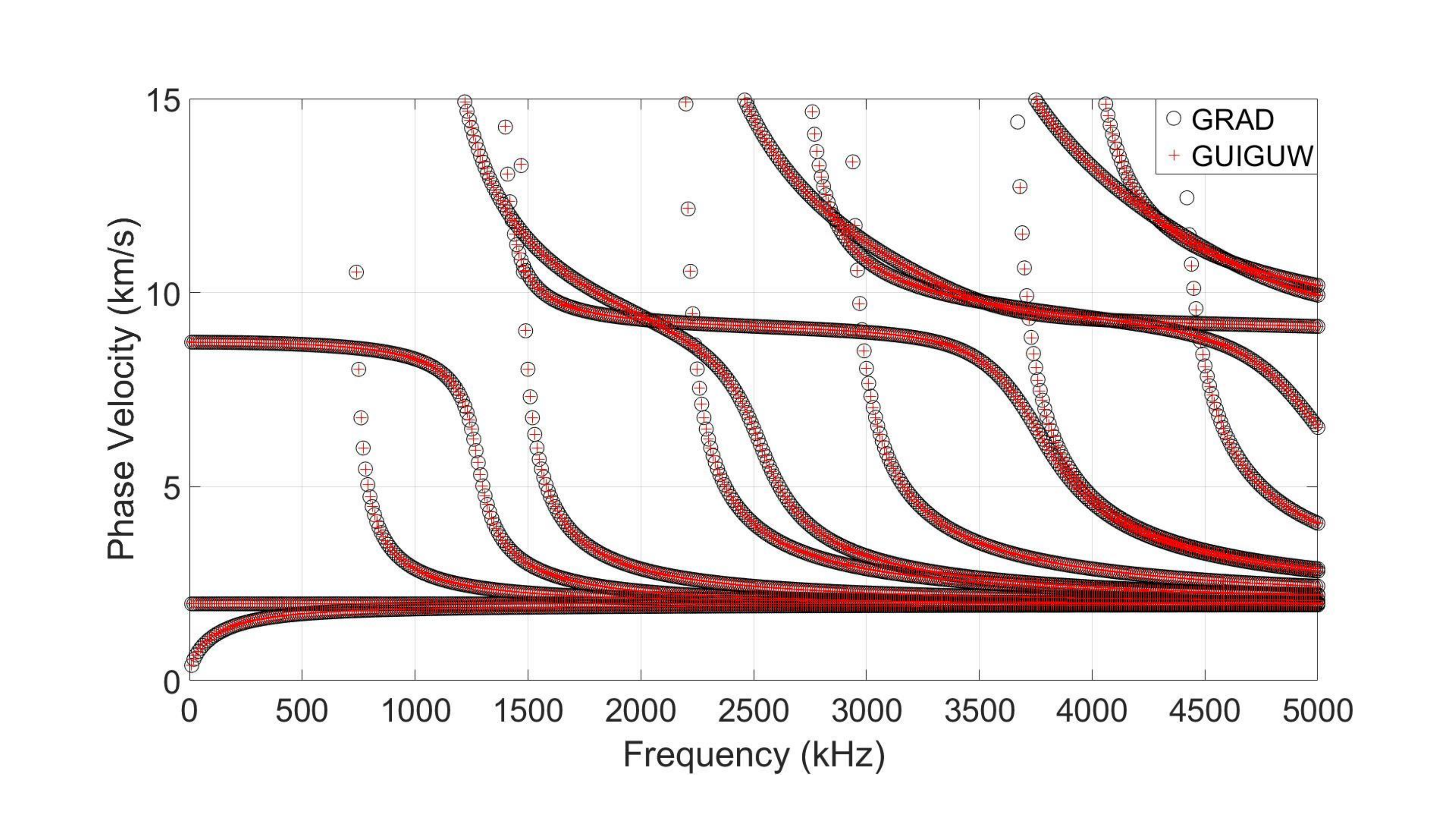}}
\subfigure[ ]{\includegraphics[trim={2.5cm 1cm 2.4cm 2cm},clip=true, scale=0.25, angle=0]{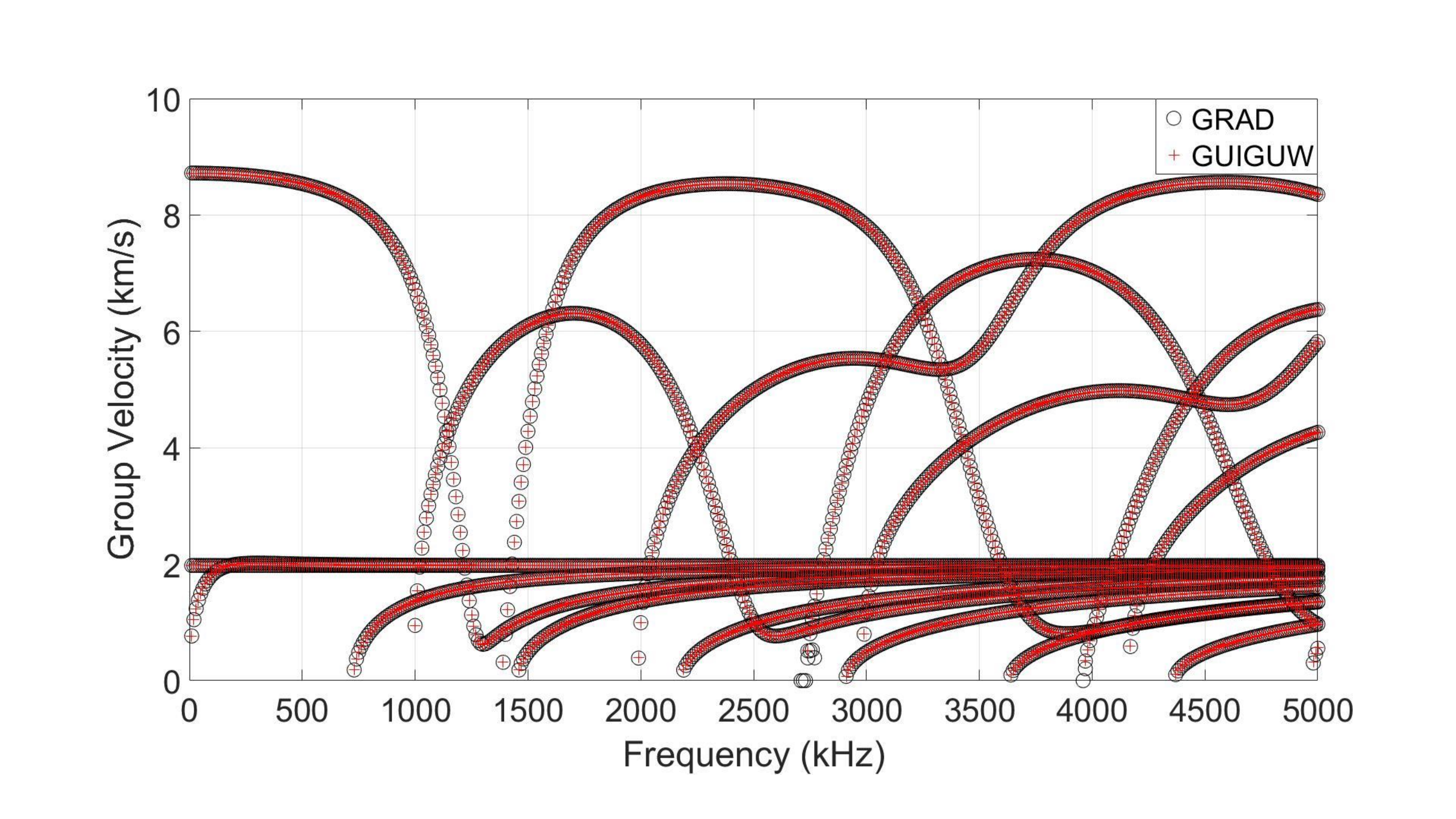}}
\subfigure[ ]{\includegraphics[trim={2.5cm 1cm 2.4cm 2cm},clip=true, scale=0.25, angle=0]{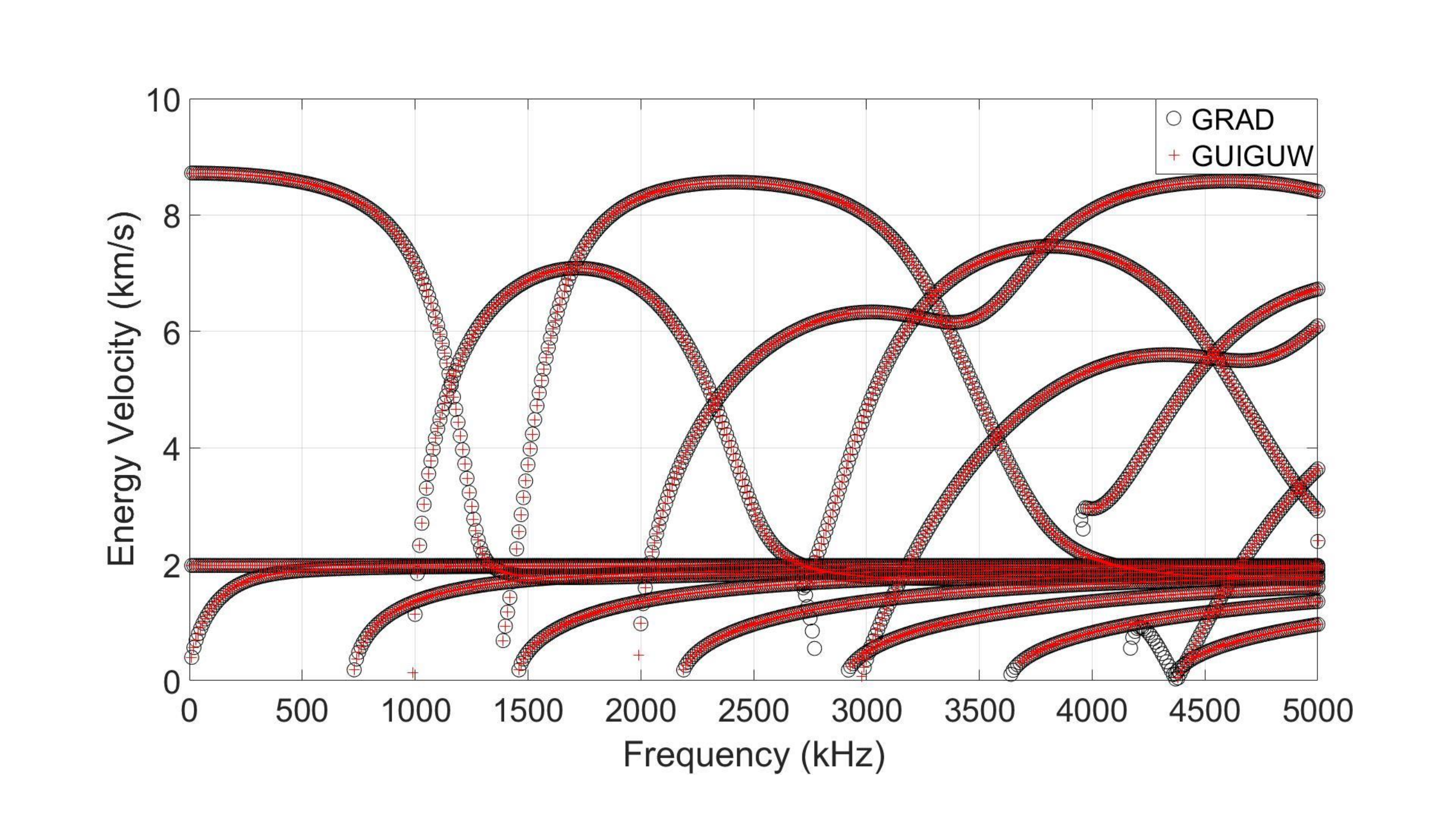}}
\caption{Comparison between GRAD and GUIGUW for C/E where $\phi=0$ (in-plane angle) (a) Undamped phase velocities, (b) Damped phase velocities, (c) Group velocities, and (d) Energy Velocities.}
\label{fig:composite}
\end{figure}

\begin{table}[htb]
\caption{Elastic and viscous properties of C/E composite lamina (in GPa).}
\begin{center}	
\begin{tabular}{cccccccccc}
\hline
$Material$ & $C_{11}$ & $C_{12}$ & $C_{13}$ & $C_{22}$ & $C_{23}$ & $C_{33}$ & $C_{44}$ & $C_{55}$ & $C_{66}$  \\ \cline{1-10}
C/E Lamina \citep{bartoli2006modeling}    & 132   & 6.9  & 12.3  & 5.9  & 5.5  & 12.1  & 3.32  & 6.21 & 6.15 \\ \cline{1-10} 
$Material$ & $\eta_{11}$ & $\eta_{12}$ & $\eta_{13}$ & $\eta_{22}$ & $\eta_{23}$ & $\eta_{33}$ & $\eta_{44}$ & $\eta_{55}$ & $\eta_{66}$  \\ \cline{1-10} 
C/E Lamina \citep{bartoli2006modeling}  & 0.4   &0.001  & 0.016  & 0.037  & 0.021  & 0.043  & 0.009  & 0.015 & 0.02  \\ \cline{1-10} 
\hline
\end{tabular}
\label{table:materials}
\end{center}		
\end{table}

The results show excellent agreement between GUIGUW and GRAD, both for the undamped and the damped case in C/E (see Fig. \ref{fig:composite}). Note that the visco-elasticity has a negligible effect on the phase velocities, whereas the difference between the group and energy velocity values is substantial, as expected. 
As the mass and stiffness matrices do not change with frequency, the use of frequency domain solutions, as well as the new compact formulaes (Eq. \ref{eq:linearized_eigen}, Eq. \ref{eq:der_eigenvalues} and Eq. \ref{eq:energy_vel}), are essential for the computational speed. In addition, the original quadratic eigenvalue problem has been reduced into a standard eigenvalue problem which requires less computational power compared to other methods. The average solution times using the proposed algorithms (average over 50 simulations) are listed in Table \ref{table:time} for computation on a workstation with Intel\textregistered Core\texttrademark i7-8700 CPU $@$ 3.20 GHz and 32 GB ram. The calculation times may change based on the computer hardware, the number of elements used in the discretization through thickness, and the number of solution points. For the most time consuming case (40 elements, 500 solution points), the solution time for GRAD only measured 112 seconds, whereas GUIGUW required 228 seconds, which demonstrates the computational efficiency of the suggested algorithms.

 \begin{table}[H]
 \caption{Averaged solution times in seconds (50 simulations).}
 \centering
 \begin{tabular}{ccccc}
 \hline
 & & \multicolumn{3}{c}{\footnotesize{$\#$ of solution points} } \\
 & & \multicolumn{1}{c}{100} & \multicolumn{1}{c}{250} & \multicolumn{1}{c}{500} \\
 \hline
 \parbox[t]{2mm}{\multirow{4}{*}{\rotatebox[origin=c]{90}{\footnotesize{$\#$ of elements}}}} &5 & 0.3373 & 0.7903  & 1.5716 \\
 &10 & 1.2028 & 3.0140  & 3.0140 \\
 &20 & 5.1893 & 13.0273 & 26.1161 \\
 &40 & 22.7285 & 55.7703 & 112.0324\\
 \hline
 \end{tabular}
 \label{table:time} 
 \end{table}

\section{Conclusion}

A novel solution strategy for the computation of eigenvalues and their derivatives in quadratic eigenvalue problems was presented. The presented strategy first linearizes the eigenvalue problem with the generalized reduction method, after which the derivatives of the eigenvalues are obtained by using dual numbers.
This method was then used to calculate Lamb wave phase, group and energy velocities in the SAFE method. 
The proposed GRAD method can also be applied to different methods, such as Legendre polynomials, higher order shear deformation theory, etc.
The introduced concepts roughly doubled the computational speed and efficiency in comparison with the semi analytical finite element method, and can be used for similar polynomial eigenvalue problems in the domain of complex wave propagation.

\section*{Acknowledgements}
The authors gratefully acknowledge the financial support from the Fund for Scientific Research-Flanders (FWO Vlaanderen, grants G066618N, G0B9515N, 1S45216N and 12T5418N), KU Leuven IF project C14/16/067 and the NVIDIA corporation.

\bibliographystyle{elsarticle-num}
\bibliography{biblio}

\end{document}